\documentclass[conference]{IEEEtran}

\usepackage{graphicx}

\usepackage{amsmath}
\usepackage{amssymb}
\usepackage{paralist}
\usepackage{graphics}
\usepackage{epsfig}
\usepackage[colorlinks=true]{hyperref}
\hypersetup{urlcolor=blue, citecolor=red}
\usepackage{algorithmic}
\usepackage{url}

\begin{document}

\title{Algorithms for High-Performance Networking in the Presence of Obstacles}

\author{\IEEEauthorblockN{Kamen Lozev}
\IEEEauthorblockA{
Email: kamen@ucla.edu}}

\maketitle

\begin{abstract}

This work develops novel algorithms for high-performance networking in the presence of obstacles based on a method for communicating via ultrasonic rays 
reflected at the obstacles. The rays are curves determined by the variable speed of sound and initial conditions and we develop ultrasonic ray 
models based on a system of differential equations. We present new parallel algorithms and software for shape and trajectory reconstruction of moving obstacles 
and show how the reconstructed reflection point of a ray at an obstacle is a natural router for
messages between the ray's transmitter and receiver and discuss the advantages of the new architecture. We discuss how the new algorithms and software improve the
performance and properties of the network architecture.

\end{abstract}

\IEEEpeerreviewmaketitle

\section{Introduction}
Consider an ultrasonic wave, or signal, described by the wave equation
\begin{equation}\label{waveequation}
u_{tt} - c^2(x,y,z)\Delta{u}=0
\end{equation}

where $c(x,y,z)>0$ is the variable speed of sound in $\Omega(t)=\Omega_0 \backslash  \overline{\Omega_1}(t)$ and

$$ u |_{\partial{\Omega_1}(t)} = 0 $$ 

for $t>0$ and $\overline{\Omega_1(t)} \subset \Omega_0 \subset \mathbb{R}^3$, where $\Omega_1(t)$ is a convex moving 
obstacle with a smooth boundary in a bounded domain $\Omega_0$.

We consider an environment without caustics and look for solutions of the form 

\begin{equation}\label{solution_wave_eq_2}
u(x,y,z,t) = \sum_{j=0}^{\infty}A_j(x,y,z) \frac{e^{i\omega(W(x,y,z)-t)}}{\omega^j}
\end{equation}

where the eikonal function $W(x,y,z)=\textit{const}$ defines a surface of constant phase. These solutions of the ray equation 
are called rays or ray solutions. Suppose that for all t we are given all integrals $\int_{\gamma} f(l) dl = C_{\gamma}$ where 
$\gamma$ are broken rays in $\Omega(t)$ such that for each point $P \in \partial{\Omega_1(t)}$ there is at least one broken ray $\gamma$ 
reflecting at P. A broken ray is a ray reflecting at the obstacle and starting and ending at the observation boundary 
$\partial{\Omega_0}$.

Let $f(x)=\frac{1}{c(x,y,z)} > 0$ in $\Omega$. Then as we know $C_{\gamma}$ correspond to signal travel times in a medium with speed of 
sound $c(x,y,z)$. The shape and trajectory reconstruction 
problem is to find $\partial{\Omega_1(t)}$ given the sets $C_{\gamma}(t)$ where $\gamma \in \Omega(t)$. 

Consider $\partial{\Omega_0}$ as the observation boundary and signal transmitters and receivers with known locations along 
the boundary. Transmitters send ray signals with known initial incident and azimuth angles at known transmission times. 
Receivers receive reflected ray signals and record the time when the signal was received. The combined data leads to a set of data points 
$$B_k=(x_l, y_l, z_l, x_r, y_r, z_r, \phi_k, \theta_k, t_k, \xi_k)$$
where  where $\phi_k$ and $\theta_k$ are the initial incident and azimuth angles of the ray from the transmitter, $x_l$, $y_l$, $z_l$ are the coordinates of the transmitter endpoint of the ray, and $x_r$, $y_r$, $z_r$ are the coordinates of 
the receiver endpoint of the ray, $t_k$ is the time of flight for the signal and $\xi_k$ is a frequency of the signal.
We present algorithms for the shape and trajectory reconstruction problem for finite travel times and variable speed of sound derived 
from the equations for the Shooting Method for two-point seismic ray tracing\cite{JG}. 

\begin{align}\label{shooting_method_equations}
\frac{dx}{dt} = c(x,y,z) \sin{\phi}\cos{\theta}\\ 
\frac{dy}{dt} = c(x,y,z) \sin{\phi}\sin{\theta}\\
\frac{dz}{dt} = c(x,y,z) \cos{\phi}\\
\frac{\partial{\phi}}{dt} = -\cos{\phi}( \frac{\partial{c}}{\partial{x}}\cos{\theta} + \frac{\partial{c}}{\partial{y}}\sin{\theta} ) + \frac{\partial{c}}{\partial{z}}\sin{\phi}\\ 
\frac{\partial{\theta}}{dt} = \frac{1}{\sin{\phi}}( \frac{\partial{c}}{\partial{x}}\sin{\theta} - \frac{\partial{c}}{\partial{y}}\cos{\theta} ) 
\end{align}

This system of equations derived from the eikonal equation has wide applications in seismology and is used in algorithms for 
seismic ray tracing \cite{SK}. The system of equations \ref{shooting_method_equations} is for a Cartesian coordinate system 
where $\vec{R}(t)=(x(t), y(t), z(t))$ is the ray position vector, $\phi(t)$ is the incident angle of the ray direction vector 
with the z axis and $\theta(t)$ is the azimuth angle that the projection of the ray direction vector makes with the positive x axis.

We consider that the speed of sound $c(x,y,z)$ is known inside $\Omega$. We will reconstruct the position of the reflection point $P$ 
given the positions of a transmitter L and receiver S, the incident and azimuth angles of the transmitted ray with respect to a Cartesian 
coordinate system centered at the transmitter, and the time of flight of the signal. We know the initial position $(x(0),y(0),z(0))$ and 
therefore we know the velocity of the ray at $t=0$. We know the angle and initial speed of wave propagation for the transmitted signal at 
time $t=0$. Then, knowing the initial conditions, we can find numerically the signal path from the transmitter L through the reflection point 
P for a given travel time $\tau_k$. Let the travel time from the transmitter L to the reflection point P be $\tau_k$ and the total travel 
time from the transmitter L to the receiver S through the reflection point P be $t_k$ where $t_k>\tau_k$.

From symmetry, we can imagine that the signal received at the receiver S is transmitted from the receiver S and that it arrives at the 
reflection point P in time $t=t_k-\tau_k$ because its travel time is $t_k-\tau_k$. The ray path from S to P is again described by the above 
system of equations for seismic ray tracing, however we do not know the initial angles for the signal from the receiver S. 

In order to reconstruct the intersection point $P_k$ of rays starting from the transmitter and receiver 
for a given data point $$B_k=(x_l, y_l, z_l, x_r, y_r, z_r, \phi_k, \theta_k, t_k, \xi_k)$$ containing the measured values for the signal, 
we step through a discrete set of values for $\tau_k$. Starting to trace a ray path from L, at each time step, each new point on the path 
is a candidate reflection point P. At this stage, this is basic initial value ray tracing. Then we apply the two point seismic ray tracing 
shooting method to see if we can reach from receiver S candidate point P for the remainder of our time budget. If we can then we have found 
the reflection point P for this data point. Otherwise, we continue with the next time step of the initial value ray tracing and check the 
next candidate point P. We repeat this until we find a reflection point P or exhaust our time budget $t_k$. 
For example, let $$\tau_{ks}=\frac{t_k n_s}{N_r}$$ where $N_r$ is an integer that specifies the time-step resolution and $n_s$ is an integer 
that specifies the number of the time step and such that $0 \leq n_s < N_r$. Each time $\tau_{ks}$  corresponds to a unique point $P_{ks}$ on 
the ray path starting from the transmitter L such that $P_{ks}$ can be reached from $L$ in time $\tau_{ks}$. Next, for each time step 
corresponding to time $\tau_{ks}$ to reach candidate $P_{ks}$ from L, we step through the range of initial angles for the signal starting 
from the receiver S and check whether the curve from S will intersect $P_{ks}$ in time $t_k-\tau_{ks}$. Alternatively, we can step through a 
range of reflection angles at the point $P_{ks}$ that is found for the given $\tau_{ks}$ and check whether the reflected signal will 
intersect S in time $t_k-\tau_{ks}$. 

\section{Reconstruction Algorithms}

The input to the following algorithm is the speed of sound $c(x)$ for the domain $\Omega$  and a set of data points or ray coordinates 
corresponding to broken rays. The output is a unique set of points in $\mathbb{R}^3$. The points from the output are the reflection points 
reconstructed from the input data.

\begin{algorithmic} \label{reconstruction_algorithm}
\REQUIRE Set of broken ray data points $B_k=(x_l, y_l, z_l, x_r, y_r, z_r, \phi_k, \theta_k, t_k, \xi_k)$
\REQUIRE Speed of sound $c(x)$ for domain $\Omega$  

\COMMENT{Algorithm for Shape and Trajectory Reconstruction of Moving Obstacles}

\COMMENT{Estimated time complexity is $O(T^2A)$ where T is the number of discretization points for the time of flight, 
and A is the number of discretization points for the angle space}           

\FORALL{data points $B_k$}

\STATE{$h_k=\frac{t_k}{N_r}$}
\STATE{$L=(X_0,Y_0,Z_0)=(x_l,y_l,z_l)$} set this initial position to be position of transmitter
\STATE{$S=(aX_0,aY_0,aZ_0)=(x_r,y_r,z_r)$} set this initial position to be position of receiver 
\STATE{$\Phi_0=\phi_k$}
\STATE{$\Theta_0=\theta_k$}
\STATE{$T_0=0$}
\STATE{$aT_0=0$}
\FOR{$s = 0 \to N_r-1$}
\STATE
\COMMENT{Compute the next point on the ray from the transmitter by fourth order Runge-Kutta step and the ray tracing system \ref{shooting_method_equations}}

\STATE{$X_{s+1} = RK4_X(h_k,T_s, X_s,Y_s,Z_s,\Phi_s,\Theta_s$)}
\STATE{$Y_{s+1} = RK4_Y(h_k,T_s, X_s,Y_s,Z_s,\Phi_s,\theta_s$)}
\STATE{$Z_{s+1} = RK4_Z(h_k,T_s, X_s,Y_s,Z_s,\Phi_s,\Theta_s$)}
\STATE{$\Phi_{s+1} = RK4_{\Phi}(h_k,T_s, X_s,Y_s,Z_s,\Phi_s,\Theta_s$)}
\STATE{$\Theta_{s+1} = RK4_{\Theta}(h_k,T_s, X_s,Y_s,Z_s,\Phi_s,\Theta_s$)}
\STATE{$T_{s+1}=T_s + h_k$}

\STATE $P_{s+1}=(X_{s+1},Y_{s+1}, Z_{s+1})$ point on solution of ray tracing equations with initial values for transmitter that is at time $T_{s+1}$ away from the transmitter L

\IF{!($P_{s+1} \in \Omega_0$)}
\STATE There must be a measurment error. Continue with next data point $B_k$
\ENDIF

\FORALL{initial angles $a\Phi_0, a\Theta_0$ in discretized angle space of the receiver}

\FOR{$p = 0 \to N_r-1$}
\STATE
\COMMENT{Compute the next point on the ray from the receiver by fourth order Runge-Kutta step and the ray tracing system \ref{shooting_method_equations}}
\STATE{$aX_{p+1} = RK4_X(h_k,aT_p,aX_p,aY_p,aZ_p,a\Phi_p,a\Theta_p$)}
\STATE{$aY_{p+1} = RK4_Y(h_k,aT_p, aX_p,aY_p,aZ_p,a\Phi_p,a\Theta_p$)}
\STATE{$aZ_{p+1} = RK4_Z(h_k,aT_p, aX_p,aY_p,aZ_p,a\Phi_p,a\Theta_p$)}
\STATE{$a\Phi_{p+1} = RK4_{\Phi}(h_k, aT_p, aX_p,aY_p,aZ_p,a\Phi_p,a\Theta_p$)}
\STATE{$a\Theta_{p+1} = RK4_{\Theta}(h_k, aT_p, aX_p,aY_p,aZ_p,a\Phi_p,a\Theta_p$)}
\STATE{$aT_{p+1}=aT_p + h_k$}

\STATE $P_{\alpha_{p+1}}=(aX_{p+1},aY_{p+1},aZ_{p+1})$ point on solution of ray tracing equations with initial angles $a\Phi_0$ and $a\Theta_0$ and initial position S, that is time $aT_{p+1}$ away from S

\IF{!($P_{\alpha_{p+1}} \in \Omega_0$)}
\STATE Exit this for loop and continue with next pair of initial angles $a\Phi_0, a\Theta_0$ from outer for loop
\ENDIF

\IF{$distance(P_{s+1},P_{\alpha_{p+1}})<\epsilon_1$ and $|T_{s+1}+aT_{p+1}-t_k|<\epsilon_2$}
\STATE $P_k=P_{s+1}$
\COMMENT{Solution for current data point $B_k$ found. Continue with next data point $B_{k+1}$}
\ENDIF

\IF{$T_{s+1}+aT_{p+1}>t_k+\epsilon_2$}
\STATE
\COMMENT{We are over the travel time budget $t_k$. Continue looking for a solution with the next set of initial angles $a\Phi_0, a\Theta_0$.}  
\ENDIF

\ENDFOR

\ENDFOR
\ENDFOR

\ENDFOR
\end{algorithmic}

RK4 stands for a fourth order Runge-Kutta method although other time-dependent numerical methods can be used as well. 
The above algorithm leads to a class of new algorithms when data structures are used to store the rays for faster processing.
For example, when rays from the receiver in its set of directions are stored and looked up in a data structure such as an array, the performance of the algorithm can be improved to $O(TA)$ where T is the number of 
discretization points for the time of flight, and A is the number of discretization points for the angle space. 
These algorithms are very suitable for parallelization and their parallelized implementation is key for efficient real-time 
processing, performance and computational efficiency.  For example, the computation for tracing the rays from the receiver 
can be parallelized in multiple threads. When the above algorithm is run on a set of points $\{B_k\}$ from one sampling time interval $T_k$, the algorithm reconstructs the shape of 
the obstacle during this sampling interval. In order to reconstruct the trajectory of the obstacle, the algorithm is run on the data points for each of 
the sampling intervals. The method can achieve high resolution because it can process a very large number of distinct points on the obstacle's 
boundary. In contrast to tomography where the focus of the reconstruction method is to recover the velocity structure of the domain, 
the shape and trajectory reconstruction procedure directly finds the shape and trajectory of the obstacle. 

\subsection{Numerical Example}
I have developed a multithreaded Java program that implements the parallelized algorithm from \ref{reconstruction_algorithm} 
using Euler's numerical integration method instead of RK4 and reconstructs the coordinates of reflection points from input data $\{B_k\}$ and 
a speed of sound function $c(x)$. Consider a circular reflecting obstacle in the plane xy moving in the first quadrant 
away from the origin along the line x=y in a medium with variable speed of sound $c(x,y)=x+y+1$. We place a transmitter and a receiver at the origin and 
set $xl = yl = zl = xr = yr = zr = 0$. In this case, the domain $\Omega_0$ is a circle of sufficiently large radius that contains the origin. 
Table \ref{Reconstruction_of_a_point_moving_on_the_line_x_y} shows the computed trajectory 
of a reflection point on the obstacle on the ray path from the origin x=y, $y>0$, corresponding to data for different travel times. 

\begin{table}[h]
\begin{center}
\begin{tabular}{|cccccc|}
	\hline
$\phi$  & $\theta$ & T & xp & yp & zp \\
  \hline
1.57 & 0.79 & 2 & 1.55 & 1.55 & 0.00 \\  
1.57 & 0.79 & 4 & 7.89 & 7.89 & 0.00 \\ 
1.57 & 0.79 & 8 & 138.15 & 138.15 & 0.00 \\ 
  \hline
\end{tabular}
\end{center}
\caption{Reconstruction of a point moving with speed $c(x,y)=x+y+1$ on the line x=y for a fixed signal frequency $\xi$.
The initial transmission angles are $\phi=\frac{\pi}{2}$ and $\theta=\frac{\pi}{4}$
\label{Reconstruction_of_a_point_moving_on_the_line_x_y}}
\end{table}

We check the accuracy of the computation in the above table by the Java program as follows. The time for the ray to reach 
to obstacle can be computed by the formula 

$$ t= \int_{0}^{X} \frac{ds}{c(s)} = \sqrt{2}\int_{0}^{X} \frac{dx}{x+y+1} = \sqrt{2}\int_{0}^{X} \frac{dx}{2x+1}$$

Therefore, 
\begin{equation}\label{numerical_example_formula}
X=Y=\frac{e^{\sqrt{2}t}-1}{2} 
\end{equation}

By symmetry, for this particular example, this time t is half of the total travel time T.
Then for a travel time $T=2$, or $t=1$, we compute $X=Y=1.55$. This result matches the corresponding result for px 
and py from Table \ref{Reconstruction_of_a_point_moving_on_the_line_x_y} obtained by numerical integration. 
In the current implementation, the reconstructed results are more accurate for shorter travel times as can be seen by comparing
the result from the table with the result from the above formula \ref{numerical_example_formula} for T=8 but this accuracy can be improved by using
numerical integration methods with a smaller error.

\subsection{Communication path from receiver to transmitter}

The reconstruction method \ref{reconstruction_algorithm} above finds the reflection point P for the signal from transmitter L to receiver S 
by finding angles of transmission $\phi$ and $\theta$ for the receiver at which a symmetric signal can be sent to L from  
S that will reach L over the reverse path of the signal detected at S. Therefore, the receiver S can communicate with
L by sending a response signal back to L along angles of transmission $\phi$ and $\theta$. L will receive this response if the 
positions of L or S have not changed signigicantly or if the speed of sound has not changed significantly in the region containing 
the communication paths. In order to create a more robust connection S can send to L several parallel rays that are very close to each other.
S and L can send and receive acknowledgments and create or use an existing networking protocol stack. This approach integrates easily with current 
network protocols and architectures. S and L recompute the reflection point and transmission angles with the receipt of each message. When
S and L receive from each other control messages with high frequency, S and L recompute a current set of transmission angles for routing back messages. 
The frequency of these control messages depends on the velocity of S, L and the obstacle, and in the case of time-varying speed of sound, 
on the time derivative of the speed of sound or on how quickly the speed of sound changes as a function of time.

\section{Network Architecture for High-Performance Communications}
 
Shape and trajectory tracking of moving obstacles that reflect ray signals sent from moving Internet hosts
enables efficient one-hop routing through reflection to the nearest Internet router or peer. A router in this 
architecture is a transmitter and receiver of ray signals. 

\subsection{Coordinate system}
The architecture is based on a known Cartesian global coordinate system. 
The physical position, for example, the high precision GPS coordinates of the transmitter or receiver 
correspond to global coordinate system coordinates $x_l, y_l, z_l$ or  $x_r, y_r, z_r$ from the
reconstruction algorithms \ref{reconstruction_algorithm}.

\subsection{Addressing}
We define a new physical layer networking protocol called IHOP. The goal of this protocol is 
to route messages through the environment in at most one hop. The IHOP address of a host is always relative 
to the position of another host, a transmitter, and consists of a pair of incident and azimuth 
angles $\phi$ and $\theta$ for reaching the receiver by sending a ray signal from transmitter to receiver. 
The IHOP protocol is defined as addressing an IHOP host from an IHOP transmitter by the host's IHOP address or by sending 
messages from the transmitter to the host along the direction determined by the IHOP angles.

\subsection{Optimal communication path}
\begin{center}
\begin{figure}
\begin{center}
\includegraphics[scale=0.40]{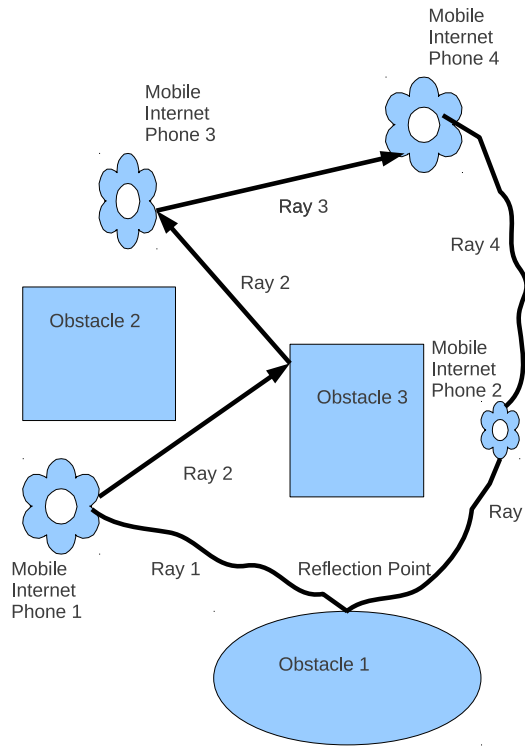}
\end{center}
\caption{Networking in the presence of obstacles. Mobile Internet Phone 1 transmits a message to Mobile Internet Phone 2 via the broken ray Ray 1. 
Knowing the speed of sound in the environment, Mobile Internet Phone 2 applies the algorithm for shape and trajectory tracking of moving obstacles from
\ref{reconstruction_algorithm} to reconstruct the reflection point at Obstacle
1 and send back response signals in the direction of this reflection point. The algorithm from \ref{reconstruction_algorithm} enables
the reconstruction of the reflection point as well as the arrival angles of the signal at Mobile Internet Phone 2.}\label{MobileInternetPhoneArchitecture}
\end{figure}
\end{center}

A transmitter communicates with a router via the optimal path, the path that takes the least time, for the given speed of sound and reflection point.
This network architecture is illustrated in Fig. \ref{MobileInternetPhoneArchitecture}. 

The combination of the methods for obstacle tracking from this work with the numerical methods for tomography 
in the presence of obstacles from \cite{L}, leads to a new method for networking with reflected 
rays based on tracking the location of reflectors and finding the speed of sound in the environment. Through reflection, 
messages can be communicated around obstacles blocking the line of sight as well as around areas with a high error rate. 
The reflectors for the shape and trajectory tracking method do not have to be installed in order to enable a network and 
can be obstacles such as buildings or cars. The two initial angles provide a way of addressing 
the obstacle and the receiver i.e. a way of calling the obstacle and the receiver by dialing two angles of transmission. 

\section{Advantages}

\subsection{Improved Security}
The communication path between transmitter and receiver is less prone to eavesdropping because communication is via a narrow ray.

\subsection{Increased Network Bandwidth}

The network architecture described in the previous section enables high bandwidth networking because a receiver can receive information
over several rays arriving from different directions. Each ray is an independent communications channel and delivers information in 
parallel with other rays detected by the receiver. 

\subsection{Increased Reliability, Survivability and Performance}
In order to increase reliability, the transmitter can send a message to the receiver along several different rays
to several different reflection points. The receiver could in this case receive the same message from several different rays or directions.
This provides a natural way for error correction and leads to high-performance networking because information from
messages that are lost or delayed could be delivered quickly from a different direction. The fastest message is processed first 
and this policy improves end-to-end performance. 
 
\subsection{Reduced delay and better support for high-performance networking}
The communication path is computed to match the speed of sound in the environment and therefore there will be less environmental errors. 
For example, the effects of temperature or humidity differences or other weather conditions are included in the computation of the communication
path via the function for the speed of sound in the environment. The reduced error rate will lead to better quality of service. The architecture is particularly
useful for streaming real-time content to home networks or peer-to-peer networks. 

\subsection{Increased Ease of Use}
The new network architecture makes it easier to connect home network devices via reflected rays compared to the current methods for wireless networking
because there is no need for additional routers. 

\section{Conclusion}

Ray signals may reach a host by an unbroken ray or by a broken ray when it reflects from an obstacle.    
For improved bandwidth and reliability, we can reach the host over different rays. The IHOP address of the host specifies a ray.

\section*{Acknowledgment}

The authors would like to thank Professor Gregory Eskin for his continuous guidance.

\newpage

\end{document}